%% file: hk-whitepaper.tex
\documentclass[12pt]{article}

\usepackage{nopageno}
\usepackage[numbers]{natbib}
\usepackage{xspace}
\usepackage{amsmath} % has \nobreakdash
\usepackage{graphicx}
\usepackage{wrapfig}

\graphicspath{{./figures/}}

\usepackage[left=1.2in,top=1.0in,bottom=1.5in,right=.9in]{geometry}

\begin{document}

% definitions here

\newcommand{\authorspace}    {\vskip 0.05in}       

\newcommand{\superk}    {Super-Kamiokande\xspace}       
\newcommand{\neutrino}  {$\nu$\xspace}
\newcommand{\nue}       {$\nu_{e}$\xspace}
\newcommand{\numu}      {$\nu_{\mu}$\xspace}
\newcommand{\nutau}     {$\nu_{\tau}$\xspace}
\newcommand{\nusterile} {$\nu_{sterile}$\xspace}
\newcommand{\mutau}     {$\nu_\mu \rightarrow \nu_{\tau}$\xspace}
\newcommand{\musterile} {$\nu_\mu \rightarrow \nu_{sterile}$\xspace}
\newcommand{\dms}       {$\Delta m^2$\xspace}
\newcommand{\sstt}      {$\sin^2 2 \theta$\xspace}
\newcommand{\numutau}     {$\nu_{\mu} \rightarrow \nu_{\tau}$\xspace}
\newcommand{\tonetwo}   {$\theta_{12}$\xspace}
\newcommand{\tonethree}   {$\theta_{13}$\xspace}
\newcommand{\ttwothree}   {$\theta_{23}$\xspace}
\newcommand{\msqonetwo}   {$\Delta m^2_{12}$\xspace}
\newcommand{\degree}      {$^\circ$\xspace}
\newcommand{\msqonethree} {$\Delta m^2_{13}$\xspace}
\newcommand{\msqtwothree} {$\Delta m^2_{23}$\xspace}
\newcommand{\pizero}{$\pi^{0} $}\newcommand{\piplus}{$\pi^{+} $}
\newcommand{\piminus}{$\pi^{-} $}
\newcommand{\Kzero}{$K^{0} $}
\newcommand{\Kplus}{$K^{+} $}
\newcommand{\eplus}{$e^{+} $}
\newcommand{\eminus}{$e^{-} $}
\newcommand{\muminus}{$\mu^{-} $}
\newcommand{\etal}{{\em et al.}}
\newcommand{\nubar}{$\overline{\nu}$\xspace}
\newcommand{\nuebar}{$\overline{\nu}_{e}$\xspace}
\newcommand{\numubar}{$\overline{\nu}_{\mu}$\xspace}
\newcommand{\nutaubar}{$\overline{\nu}_{\tau}$\xspace}
\newcommand{\pe}{${\it p.e.}$}
\newcommand{\ie}{${\it i.e.}$}
\newcommand{\Ec}{$E_{C}$}
\newcommand{\SK}{Super-K\xspace}
\newcommand{\HK}{Hyper-K\xspace}

\title{\bf Hyper-Kamiokande Physics Opportunities}
\date{August 31st 2013}
\author{\Large Submitted by the Hyper-Kamiokande Working
  Group~\footnote{Project contact: Tsuyoshi Nakaya $<$t.nakaya@scphys.kyoto-u.ac.jp$>$} \\
  to the 2013 Snowmass Process}

\maketitle

\begin{abstract}

  We propose the Hyper-Kamiokande (Hyper-K) detector as a next
  generation underground water Cherenkov
  detector~\cite{Abe:2011ts}. It will serve as a far detector of a
  long baseline neutrino oscillation experiment envisioned for the
  upgraded J-PARC beam, and as a detector capable of observing, far
  beyond the sensitivity of the Super-Kamiokande (Super-K) detector,
  proton decays, atmospheric neutrinos, and neutrinos from
  astrophysical origins. The current baseline design of Hyper-K is
  based on the highly successful Super-K detector, taking full
  advantage of a well-proven technology.  Hyper-K consists of two
  cylindrical tanks lying side-by-side, the outer dimensions of each
  tank being $\textrm 48 (W) \times 54 (H) \times 250 (L) ~ m^3$. The
  total (fiducial) mass of the detector is 0.99 (0.56) million metric
  tons, which is about 20 (25) times larger than that of Super-K. A
  proposed location for Hyper-K is about 8 km south of Super-K (and
  295 km away from J-PARC) at an underground depth of 1,750 meters
  water equivalent (m.w.e.). The inner detector region of the Hyper-K
  detector is viewed by 99,000 20-inch PMTs, corresponding to the PMT
  density of 20\% photo-cathode coverage (one half of that of
  Super-K).

  The Hyper-K project is envisioned to be completely open to the
  international community.  The current working group contains members
  from Canada, Japan, Korea, Spain, Switzerland, Russia, the United Kingdom
  and the United States. 
  The United States physics community has a long history of making
  contributions to the neutrino physics program in Japan. In
  Kamiokande, Super-Kamiokande, K2K and T2K, US physicists have played
  important roles building and operating beams, near detectors, and
  large underground water Cherenkov detectors.
  This set of three one-page whitepapers prepared for the US Snowmass
  process describes the opportunities for future physics discoveries
  at the Hyper-K facility with beam, atmospheric and astrophysical
  neutrinos.

\end{abstract}

%\begin{figure}[!h]
 % \centering    
 % \includegraphics[width=2.0in]{HK-SchematicView_5}
%\end{figure}

\newpage
{\small \input{author.tex} }            \newpage
\input{beam-physics.tex} \newpage
\input{atmospheric.tex}    \newpage
\input{low-energy.tex}     \newpage
%\input{pdk.tex}                 \newpage

%%%
%%% We are using bibtex
%%%
\bibliographystyle{apsrev}
\bibliography{references}

\end{document}

%% file: author.tex
\begin{center}
\vspace{0.0in} \textbf{\Large The Hyper-Kamiokande Working Group} \vspace{.15in}
\end{center}

\noindent
{\bf Boston University (USA):} %Contact: E. Kearns \authorspace
E. Kearns,  J.L. Stone \authorspace

\noindent
{\bf Chonnam National University (Korea):} %Contact: K.K. Joo \authorspace
K.K. Joo \authorspace

\noindent
{\bf Duke University (USA):} %Contact: C.W. Walter \authorspace
T. Akiri, A. Himmel,  K. Scholberg, C.W. Walter \authorspace

\noindent 
{\bf Earthquake Research Institute, The University of Tokyo (Japan):} %Contact: A. Taketa \authorspace
 A. Taketa,  H.K.M. Tanaka \authorspace

\noindent
{\bf ETH Zurich (Switzerland):} %Contact: A. Rubbia \authorspace
 A. Rubbia \authorspace

\noindent
{\bf Institute for Nuclear Research (Russia):} %Contact: Y. Kudenko  \authorspace
A. Izmaylov, M. Khabibullin,  Y. Kudenko  \authorspace

\noindent
{\bf Imperial College London (UK):}
 M. Malek, Y. Uchida, M.O. Wascko \authorspace

\noindent
{\bf Iowa State University (USA):} 
I. Anghel, G. Davies,  M.C. Sanchez, T. Xin \authorspace

\noindent 
{\bf Kamioka Observatory, ICRR, The University of Tokyo (Japan):} %Contact: M. Shiozawa \authorspace
K. Abe, Y. Haga, Y. Hayato, J. Kameda, Y. Kishimoto,
 M. Miura, S. Moriyama, M. Nakahata, S. Nakayama, H. Sekiya,
M. Shiozawa, Y. Suzuki, A. Takeda, H. Tanaka, T. Tomura, R. Wendell \authorspace

\noindent
{\bf Kavli IPMU, The University of Tokyo (Japan):}  %Contact: M.R. Vagins  \authorspace
M. Hartz, L. Marti, K.Nakamura, M.R. Vagins  \authorspace

\noindent
{\bf KEK (Japan):} %Contact: T. Kobayashi \authorspace
M. Friend, T. Ishida, T. Kobayashi, Y. Oyama \authorspace

\noindent
{\bf Kobe University (Japan):} %Contact: Y. Takeuchi \authorspace
 A. T. Suzuki, Y. Takeuchi \authorspace

\noindent
{\bf Kyoto University (Japan):} %Contact: T. Nakaya \authorspace
S. Hirota, K. Huang, A. K. Ichikawa, M. Ikeda, A. Minamino, T. Nakaya,
K. Tateishi \authorspace

\noindent
{\bf Lancaster University (UK):}
A. Finch, L.L. Kormos,  J. Nowak, H.M. O'Keeffe, P.N. Ratoff

\noindent
{\bf Los Alamos National Laboratory (USA):} %Contact: G. Sinnis \authorspace
%C. Mauger, 
G. Sinnis \authorspace

\noindent 
{\bf Louisiana State University (USA):} %Contact: T. Kutter \authorspace 
F.d.M. Blaszczyk,  J. Insler, T. Kutter, O. Perevozchikov, M. Tzanov \authorspace

\noindent
{\bf Miyagi University of Education (Japan):} %Contact: Y.Fukuda \authorspace
Y.Fukuda \authorspace

\noindent
{\bf Nagoya University (Japan):} %Contact: Y. Itow \authorspace
K.Choi, T. Iijima,  Y. Itow \authorspace

%\noindent
%{\bf The Ohio State University (USA):} Contact: C. Rott \authorspace

\noindent
{\bf Okayama University (Japan):} %Contact: Y. Koshio \authorspace
H. Ishino, Y. Koshio, T. Mori, M. Sakuda, T. Yano \authorspace

\noindent
{\bf Osaka City University (Japan):} %Contact: K. Yamamoto \authorspace
Y. Seiya,  K. Yamamoto \authorspace

\noindent 
{\bf Pontif{\'\i}cia Universidade Cat{\'o}lica do Rio de Janeiro (Brazil):} %Contact: H. Nunokawa \authorspace
H. Minakata, H. Nunokawa \authorspace

\noindent
{\bf Queen Mary, University of London (UK):}
F. Di Lodovico, T. Katori, R. Sacco, B. Still, R. Terri, J.R. Wilson \authorspace

\noindent
{\bf Seoul National University (Korea):} %Contact: S. B. Kim  \authorspace
S. B. Kim  \authorspace

\noindent
{\bf State University of New York at Stony Brook (USA):}
J. Adam, J. Imber, C. K. Jung, C. McGrew, J.L. Palomino, C. Yanagisawa \authorspace

\noindent
{\bf STFC Rutherford Appleton Laboratory (UK):}
D. Wark, A. Weber \authorspace

\noindent
{\bf Sungkyunkwan University (Korea):} %Contact C.~Rott \authorspace
C.~Rott \authorspace

\noindent
{\bf The California State University Dominguez Hills (USA)}: %Contact K. Ganezer \authorspace
K. Ganezer, B. Hartfiel, J. Hill

\noindent
{\bf The University of Tokyo (Japan):} %Contact: M. Yokoyama \authorspace
H. Aihara, Y. Suda, M. Yokoyama \authorspace

\noindent
{\bf Tohoku University (Japan):} %Contact: K. Inoue \authorspace
K. Inoue,  M. Koga, I. Shimizu \authorspace

\noindent
{\bf Research Center for Cosmic Neutrinos, ICRR, The University of
  Tokyo (Japan):}  %Contact: K.  Okumura \authorspace
T. Irvine, T. Kajita, I. Kametani, Y. Nishimura, K. Okumura, E. Richard \authorspace

\noindent
{\bf Tokyo Institute of Technology (Japan):} %Contact: M. Kuze \authorspace
M. Ishitsuka, M. Kuze, Y. Okajima \authorspace

\noindent
 {\bf TRIUMF (Canada):} %Contact: A. Konaka \authorspace
P. Gumplinger, A. Konaka, T. Lindner, K. Mahn, J.-M. Poutissou, F.
Retiere, M. Scott, M.J. Wilking \authorspace

\noindent
{\bf University Autonoma Madrid (Spain):}
L. Labarga \authorspace

\noindent
{\bf University of British Columbia (Canada):} %Contact: H.A. Tanaka \authorspace
S.M. Oser, H.A. Tanaka \authorspace

\noindent
{\bf University of Geneva (Switzerland):}% Contact: A. Blondel \authorspace
A. Blondel, A. Bravar, F. Dufour, Y. Karadhzov, A. Korzenev, E. Noah,
M. Ravonel, M. Rayner, R. Asfandiyarov, A. Haesler, C. Martin, E. Scantamburlo \authorspace

\noindent
{\bf University of Hawaii (USA):} %Contact: J.G. Learned \authorspace
 J.G. Learned \authorspace

\noindent
{\bf University of Regina (Canada):} %Contact: M. Barbi \authorspace
M. Barbi \authorspace

\noindent
{\bf University of Toronto (Canada):} %J.F. Martin \authorspace
J.F. Martin \authorspace

\noindent
{\bf University of California, Davis (USA):} %Contact: R.Svoboda \authorspace
M. Askins, M.Bergevin, R. Svoboda \authorspace

\noindent
{\bf University of California, Irvine (USA):} %Contact: H.W. Sobel \authorspace
G. Carminati, S. Horiuchi, W.R. Kropp, S. Mine, M.B. Smy, H.W. Sobel \authorspace

\noindent
{\bf University of Liverpool (UK):}
N. McCauley,  C. Touramanis \authorspace

\noindent
{\bf University of Oxford (UK):}
G. Barr,  D. Wark, A. Weber \authorspace

\noindent
{\bf University of Pittsburgh (USA):} %Contact: V. Paolone\authorspace
V. Paolone \authorspace

\noindent
{\bf University of Sheffield (UK):}
J.D. Perkin, L.F. Thompson \authorspace

\noindent
{\bf University of Washington (USA):} %Contact: R.J. Wilkes \authorspace
J. Detwiler, N. Tolich, R. J. Wilkes \authorspace

\noindent
{\bf University of Warwick (UK):}
G.J. Barker, S. Boyd, D.R. Hadley, M.D. Haigh \authorspace

\noindent
{\bf University of Winnipeg (Canada):} 
B. Jamieson \authorspace

\noindent
{\bf Virginia Tech (USA):} %Contact: C. Mariani \authorspace
S. M. Manecki, C. Mariani, S. D. Rountree, R. B. Vogelaar \authorspace

\noindent
{\bf York University (Canada):} %Contact:  S. Bhadra \authorspace
S. Bhadra \authorspace

%%% Local Variables: 
%%% mode: latex
%%% TeX-master: "hk-whitepaper.tex"
%%% End: 

%% file: beam-physics.tex
\noindent
{\bf \large Exploring CP violation with the upgraded
  J-PARC Beam}
{\vskip 0.075in}

\begin{wrapfigure}{r}[.05cm]{0in}
  \centering
  \includegraphics[width=2.5in,angle=90]{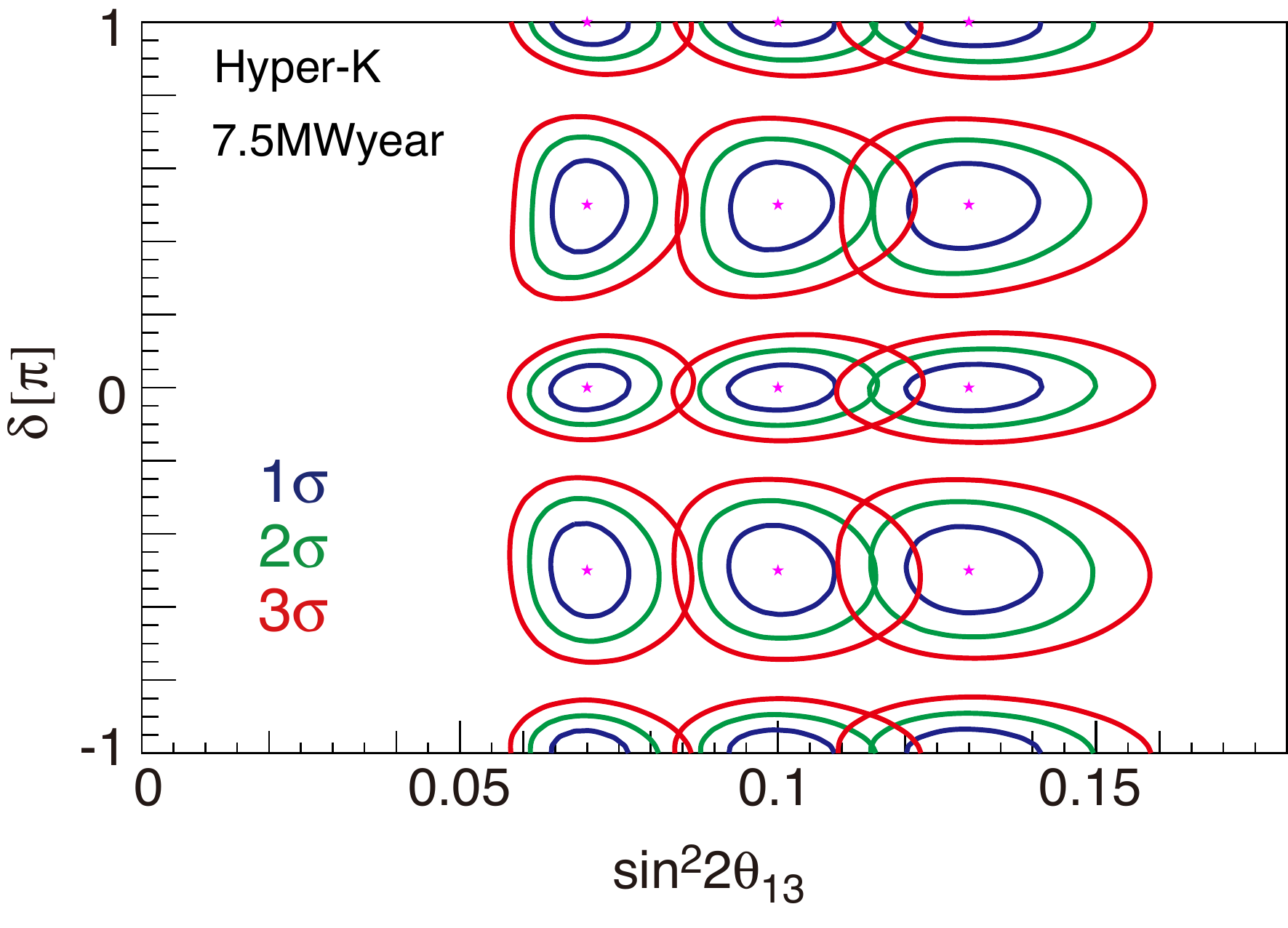}
  \caption{Allowed regions in the space of $\sin^22\theta_{13}$ and
    $\delta$ near the known value of $\sin^22\theta_{13}$.  Blue,
    green, and red lines represent 1, 2, 3 $\sigma$ allowed regions in
    with the hierarchy known to be normal.}
\label{fig:jellybean}
\end{wrapfigure}

In 2011, the T2K~\cite{Abe:2011sj}, MINOS~\cite{Adamson:2011qu}, and
Double Chooz~\cite{Abe:2011fz} experiments showed the first
indications of full three-flavor oscillations.  In 2012 the Daya
Bay~\cite{An:2012eh} and RENO~\cite{Ahn:2012nd} experiments reported
the first precision measurements of the $\theta_{13}$ mixing angle
which drives three-flavor oscillation.  This unexpectedly large value
of \tonethree guarantees the ability to measure the CP violating
phase $\delta$.  The Hyper-K projects builds on the proven success of
the water Cherenkov technique with an upgraded J-PARC beam plus a
megaton-scale detector.

The J-PARC to Hyper-K experiment will use a proton beam with a
$\sim \textrm 1 MW$ of power to produce both neutrinos and
anti-neutrinos with a horn focusing system. The beam is directed
2.5$^\circ$ off-axis to the 560~kton fiducial volume Hyper-K detector
295~km away.  The neutrinos are measured near the production site and
then again at the far detector.  $CP$ violation will manifest itself
in a difference between the measured rate of $\nu_\mu \rightarrow
\nu_e$ and $\bar\nu_\mu \rightarrow \bar\nu_e$ transformations as well
as in spectral distortions. The high event rate in the far detector
will also allow for precise measurements of neutrino mixing parameters
that will continue the worldwide effort to constrain the $3 \times 3$
PMNS matrix.

\begin{wrapfigure}[13]{l}[.05cm]{0in}
  \centering
  \includegraphics[width=2.75in]{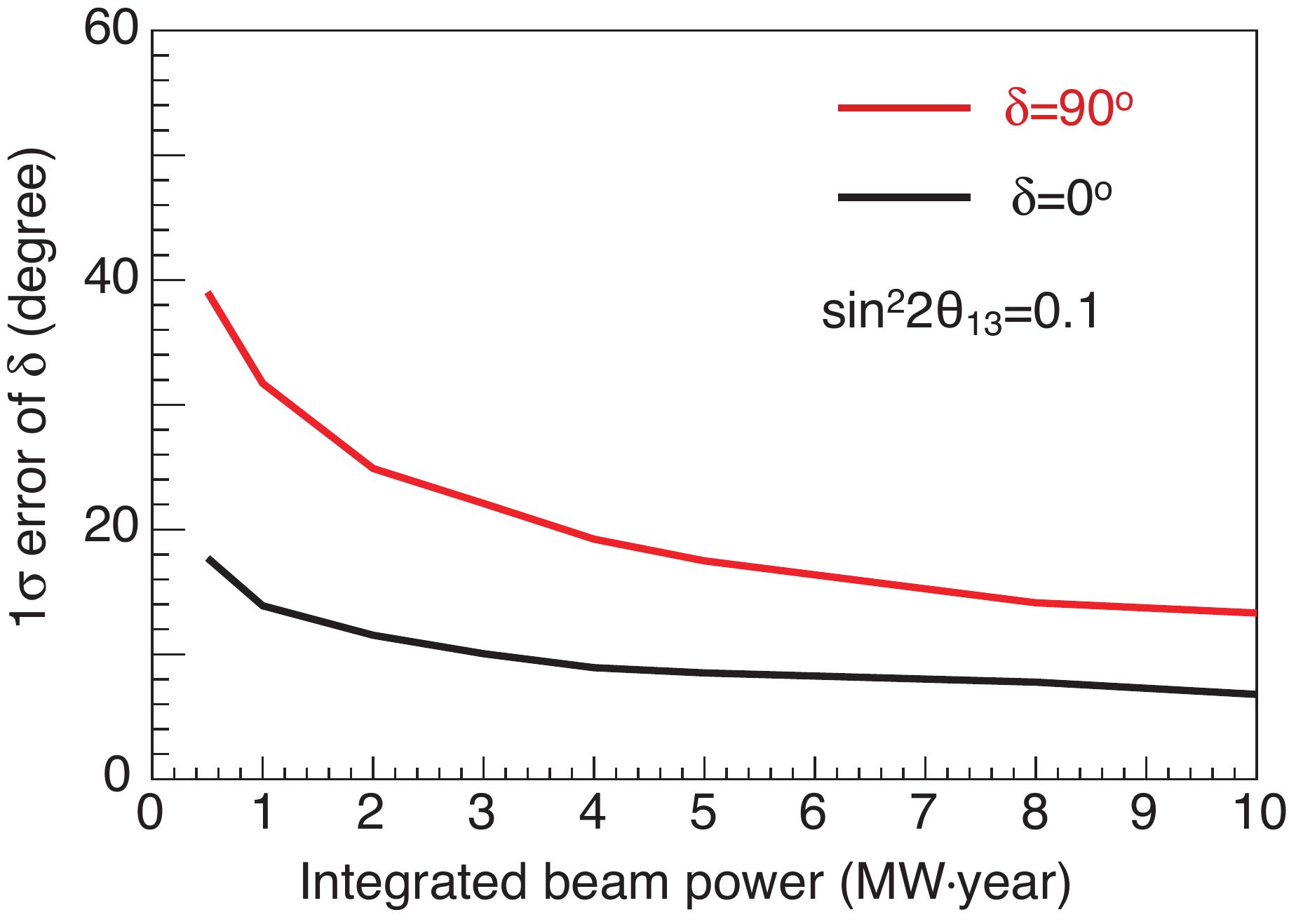}
  \caption{1$\sigma$ uncertainty of $\delta$ as a function of
    integrated beam power.  The ratio of $\nu$ and $\bar{\nu}$ running
    time is fixed to 3:7.}
  \label{fig:delta-width}
\end{wrapfigure}

\noindent
In 10 years of running, split 3:7 between neutrinos and antineutrinos
respectively, we expect to collect approximately 2000 to 4000 signal
events in each mode. Assuming the mass hierarchy is determined by
other means, and that the total systematic error managed to the 5\%
level, then the CP phase may be distinguished from $\delta = 0$ at
3$\sigma$ for 74\% of the entire range in
$\delta$. Figure~\ref{fig:jellybean} shows example 1$\sigma$ contours
using the current range of $\sin^2 2
\theta_{13}$. Figure~\ref{fig:delta-width} demonstrates the expected
uncertainty in $\delta$ as a function of integrated beam power for two
values of $\delta$.  After 7.5 MW$\cdot$years, $\delta$ will be
measured with an accuracy between 7 and 15 degrees.

%%% Local Variables: 
%%% mode: latex
%%% TeX-master: "hk-whitepaper.tex"
%%% End: 

%% file: atmospheric.tex
\noindent
{\bf \large Exploring Neutrino Properties with 
Atmospheric Neutrinos}
{\vskip 0.075in}

In the late nineties, atmospheric neutrinos measured in the
Super-Kamiokande detector provided the first definitive evidence that
neutrinos had mass and that the mass states mixed to make the well
known flavor states~\cite{fukuda:1998mi}. Atmospheric neutrinos remain
an important probe of neutrino oscillations, and the large statistics
sample from the one-half megaton Hyper-K will offer an unprecedented
opportunity to study them in detail.  Atmospheric neutrinos exist in
both neutrino and anti-neutrino varieties in both muon and electron
flavors.  Approximately 1,000,000 events are expected to be collected
in a 10 year period.
The large value of \tonethree, along with the neutrino versus
anti-neutrino dependent matter resonance effect in the earth opens up
the study of oscillation driven electron neutrino appearance.  The
oscillation effect in the electron neutrino flux have been
analytically calculated~\cite{Peres:2003wd} as:

\begin{eqnarray}
 \frac{\Phi(\nu_e)}{\Phi_0(\nu_e)} -1 & \approx & P_2\cdot(r\cdot \cos^2\theta_{23
}-1) \nonumber \\
&& -r\cdot \sin\tilde{\theta}_{13}\cdot \cos^2\tilde{\theta}_{13}\cdot \sin{2\theta_{23}}\cdot(\cos{\delta}\cdot R_2-\sin{\delta}\cdot I_2) \nonumber \\
&& +2\sin^2\tilde{\theta}_{13}\cdot(r\cdot \sin^2\theta_{23}-1)
\label{eqn:nue-oscillation}
\end{eqnarray}

\begin{wrapfigure}{r}[.05cm]{0in}
  \includegraphics[width=2.25in]{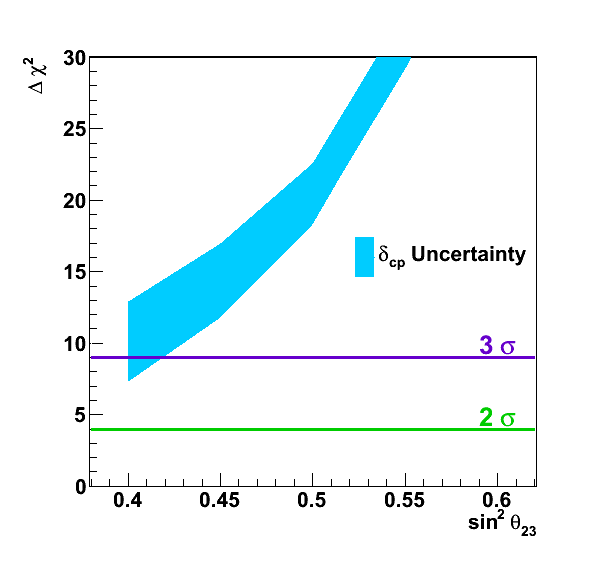}
  \caption{The sensitivity to the mass hierarchy
    as a function of
    \ttwothree with \tonethree fixed at $\sin^2 2 \theta_{13} = 0.098$
    (NH case).}
   \label{fig:MH}
\end{wrapfigure}

{\vskip 0.025in}
\noindent
where we call the first, second, and third terms the ``solar term'',
``interference term'', and ``$\theta_{13}$ resonance term'',
respectively.  \(P_2\) is the two neutrino transition probability of
\(\nu_e \rightarrow \nu_{\mu,\tau}\) which is driven by the solar
neutrino mass difference \(\Delta m^2_{21}\).  $R_2$ and $I_2$
represent oscillation amplitudes for $CP$ even and odd terms.  For
anti-neutrinos, the sign of the $\delta$ should be changed.
Additionally, the modified probabilities for $P_2, R_2, I_2$ are
obtained by replacing the matter potential $V \rightarrow -V$ (see
\cite{Peres:2003wd} for details).  The electron appearance effect
along with precision measurements of muon
disappearance~\cite{ashie:2005ik} and tau appearance~\cite{Abe:2012jj}
will allow Hyper-K to probe the octant of \ttwothree oscillation, the
mass hierarchy and CP violation phase.

A full Monte Carlo and reconstruction study using Super-Kamiokande
tools has determined that the expected significance for the mass
hierarchy determination is more than $3 \sigma$ provided $\sin^2
\theta_{23} > 0.4$. We expect to be able to discriminate between
$\sin^2 \theta_{23} < 0.5$ (first octant) and $> 0.5$ (second octant)
at the $3 \sigma$ level if $\sin^22 \theta_{23}$ is less than 0.99.
For all values of $\delta$, 40\% of the $\delta$ range can be
excluded at three sigma assuming that $\sin^2 \theta_{23} > 0.4$.  All
of these results are obtained using atmospheric neutrinos alone.  In
combination with the JPARC beam they can be even more tightly
constrained.  As an example, figure~\ref{fig:MH} demonstrates the
sensitivity to the mass hierarchy as a function of \ttwothree with
\tonethree fixed at $\sin^2 2 \theta_{13} = 0.098$ for the case of the
normal hierarchy.

%%% Local Variables: 
%%% mode: latex
%%% TeX-master: "hk-whitepaper.tex"
%%% End: 

%% file: low-energy.tex
\noindent
{\bf \large Low-Energy Neutrino Physics and Astrophysics with Hyper-K}
{\vskip 0.075in}

\noindent
Hyper-K represents the next generation of highly-successful water
Cherenkov technology for observation of low-energy (less than $\sim$50
MeV) neutrinos, including solar
neutrinos~\cite{Hosaka:2005um,Cravens:2008aa,Abe:2010hy},
core-collapse supernova
neutrinos~\cite{Hirata:1987hu,Ikeda:2007sa,Bays:2011si}, and
potentially dark matter annihilation neutrinos~\cite{Rott:2012qb}.

{\vskip 0.05in}
\noindent
{\bf Solar Neutrinos:}
Assuming sufficient depth to overcome cosmic-muon-induced spallation
background, Hyper-K will observe $\sim$115,000 elastic solar $^8$B
neutrino-electron scatters per year (above its energy threshold and
after event selection efficiency, assuming a livetime of 90\%), an
unprecedentedly large solar neutrino rate.  With tight control of
systematic uncertainties, 
Hyper-K will be very sensitive to the solar
day/night effect, \textit{i.e.} the regeneration of $\nu_e$ flavor of
solar neutrinos passing through the Earth. Within five years, Hyper-K
will determine the solar zenith angle variation amplitude to
$\sim$0.5\%-- expressed here as a day/night asymmetry, defined as
$({\rm day}-{\rm night})/(0.5({\rm day}+{\rm night}))$-- measuring
$\Delta m^2_{21}$ with a precision comparable to that of current
reactor antineutrino experiments, and establishing (or refuting) the
presence of matter effects on neutrino oscillations with a
significance exceeding 4$\sigma$~\cite{nu2012}.

\begin{wrapfigure}{r}[.05cm]{0in}
  \includegraphics[width=2.25in]{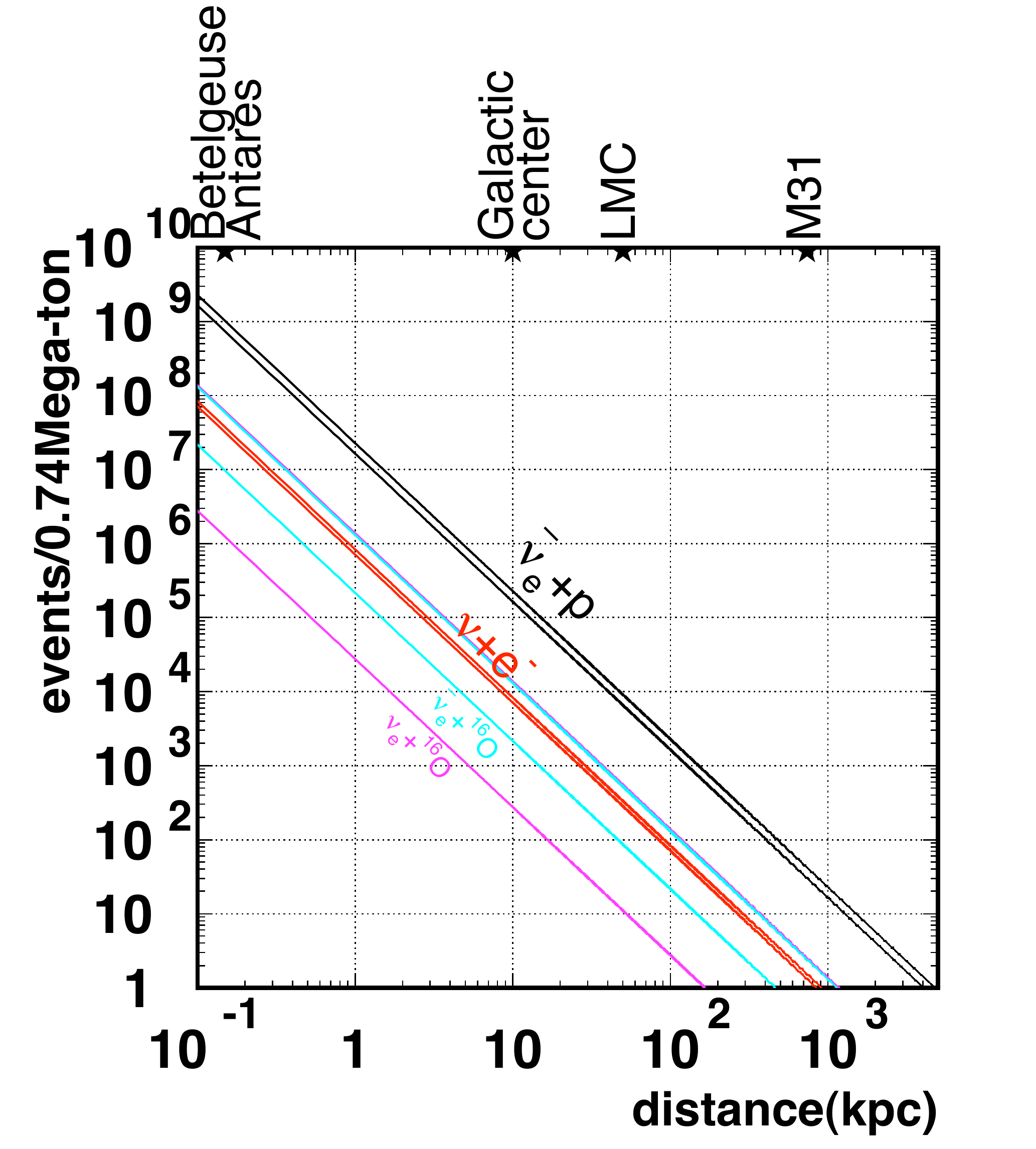}
  \caption{Expected numbers of SN burst events in HK for each
    interaction as a function of the distance to the
    SN~\cite{Abe:2011ts}. }
 \label{fig:SN}
\end{wrapfigure}

{\vskip 0.05in}
\noindent
{\bf Supernova Neutrinos:}
Hyper-K is sensitive to neutrinos from Galactic core-collapse
supernovae ($\sim$250,000 interactions in $\sim$10 s at the Galactic
center a few times/century), nearby supernovae ($\sim$25 interactions
at Andromeda; 1/2 interactions for distances $<4$~MPc with $\sim 60\%/
25\%$ probability every few years)
%\footnote{There have been more than five such supernovae in
%the last decade.}) ]
and distant supernovae ($\sim$100 interactions/year, up to $z\sim1$).
With Gd salt doping~\cite{Beacom:2003nk}, Hyper-K could also separate $\bar{\nu}_e$ inverse beta
reactions 
%from elastic scattering (mostly $\nu_e$) and 
from other interactions.  A large-statistics Galactic burst offers many unique
opportunities (\textit{e.g.} \cite{Dighe:2008dq,Scholberg:2012id} and
references therein).  From the time, flavor and energy profile of the neutrinos, we
can learn about the neutronization burst, shock wave effects, explosion
temperatures, and black hole formation.  
We may gain information on neutrino parameters, in
particular mass hierarchy.  Spectral swaps between neutrino flavors
will enable the study of $\nu-\nu$ interactions.  In addition to an
early alert for even quite distant supernovae, Hyper-K could also
achieve precision pointing to a nearby supernova's direction.
% a few hours before the arrival of the light. 
%Despite low statistics, 
%nearby
%extragalactic supernova neutrinos will tell us about differences
%between supernova explosions; 
Even a few neutrinos from nearby extragalactic supernovae will determine the nature
of nearby transients whose mechanism is uncertain~\cite{Thompson:2008sv}.
%whether they are truly core-collapse supernovae or perhaps an entirely
%different phenomenon can be settled by the detection of even a few
%neutrinos.
 The rate and spectrum of distant
supernova neutrinos characterize the properties of typical supernova
explosions (luminosity and explosion temperature).  For distant
supernovae, Gd doping is essential to tag $\bar{\nu}_e$
between 10 and 30 MeV and thereby distinguish supernova neutrinos from
atmospheric neutrino backgrounds.  
% (such as $\nu_\mu$
%charged-current interactions for which the muon is below Cherenkov
%threshold).

% \begin{figure}[!htb]
%   \centering
%     \begin{minipage}{3.in}
%       \centering
%       \includegraphics[width=1.5in]{sn-spectrum.pdf}
%     \end{minipage}
%     \begin{minipage}{3.in}
%       \centering
%       \vspace{0.75in}
%       \includegraphics[width=1.5in]{sn-evtvsdist.pdf}
%     \end{minipage}
%     \caption{Left:  Right:}
%     \label{fig:nue-plots}
% \end{figure}

%%% Local Variables: 
%%% mode: latex
%%% TeX-master: "hk-whitepaper.tex"
%%% End: 